\documentclass[12pt]{article}
\usepackage{graphicx,amssymb, amsmath}
\newcommand{\mathsym}[1]{{}}

\let\badcite=\cite
\def\cite{~\badcite}

\def\slashchar#1{\setbox0=\hbox{$#1$}           
   \dimen0=\wd0                                 
   \setbox1=\hbox{/} \dimen1=\wd1               
   \ifdim\dimen0>\dimen1                        
      \rlap{\hbox to \dimen0{\hfil/\hfil}}      
      #1                                        
   \else                                        
      \rlap{\hbox to \dimen1{\hfil$#1$\hfil}}   
      /                                         
   \fi}
    \def\slashword#1{\setbox0=\hbox{$#1$}        
  \dimen0=\wd0                                   
   \setbox1=\hbox{/} \dimen1=\wd1                
   \ifdim\dimen0>\dimen1                         
      \rlap{\hbox to \dimen0{\hfil\bf---\hfil}} %
      #1                                         %
   \else                                         
      \rlap{\hbox to \dimen1{\hfil$#1$\hfil}}    
      /                                          
    \fi}                                         %

\catcode`@=11
\newdimen\vbigd@men                             

\def\vbig#1#2{{\vbigd@men=#2\divide\vbigd@men by 2%
   \hbox{$\left#1\vbox to \vbigd@men{}\right.\n@space$}}}
\catcode`@=12

\catcode`@=11
\def\citenum#1{\csname b@#1\endcsname}
\catcode`@=12

\begin{document}
\begin{titlepage}

\bigskip\bigskip

\begin{center}{\Large\bf\boldmath
Fast and Fuzzy Private Set Intersection
}
\end{center}
\bigskip
\centerline{\bf Nicholas Kersting
\footnote{Email: 1054h34@gmail.com}
 }

\bigskip

\begin{abstract}
Private Set Intersection (PSI) is usually implemented as a sequence of encryption rounds between pairs of users, whereas the present work implements PSI in a simpler fashion: each set only needs to be encrypted once, after which each pair of users need only one ordinary set comparison. This is typically orders of magnitude faster than ordinary PSI at the cost of some ``fuzziness" in the matching, which may nonetheless be tolerable or even desirable. This is demonstrated in the case where the sets consist of English words processed with WordNet.
\end{abstract}

\newpage
\pagestyle{empty}

\end{titlepage}


\section{Introduction}
\label{sec:introduction}
Private Set Intersection (PSI) is a process wherein two or more parties compare sets to mutually discover their intersection without revealing anything else about the sets\cite{psi1,psi2,psi3,psi4}. Though important potential applications abound in e.g. medicine, finance, and networking, a convincingly practical realization of PSI has yet to surface. State-of-the-art techniques employ a series of interaction rounds between users with compound encryptions of their sets\cite{psi5, psi6}, typically taking $O(10^2)$ milliseconds to compare small sets consisting of a few hundred elements. This situation is particularly acute in the context of smartphone apps, where we have probably the most exciting and relevant use cases of PSI\cite{smart1, smart2}, but the severe CPU and memory constraints there slow the performance of existing PSI algorithms to seconds or minutes.

 This paper introduces an algorithm to implement PSI in a different way: sets are first encrypted and then compared in one round as ordinary sets. This promises to be orders of magnitude faster than ordinary PSI in situations where set content is unvarying over many comparisons, e.g. as would occur in social matchmaking applications, for then the cost of encryption is amortized over the many comparison operations. Since the encryption only needs to be done once on the input set after which it can be compared any number of times to other sets, these encryptions enjoy the ability to be stored, ported, or subjected to additional layers of security as desired. Moreover, the technique is naturally suited to sets of extended elements, for example English words, where intersection is not best expressed as a binary ``yes/no" but rather as a continuous score between 0 and 1.

 In this algorithm speed is attained at the cost of an inherent uncertainty or ``fuzziness" in the comparison process: the resultant intersection between the sets will generally contain false positives (but no false negatives), however these can be reduced to a tolerable level by methods privy to the particular application at hand. At the same time, this fuzziness actually assists as a guard against brute-force decryption, which in addition to being NP-hard thus also produces too many irreducible false positives to be informational. Fuzziness is therefore more of a feature than a liability of the algorithm.

The following exposition will define the encryption technique, apply it to an intuitive geometrical toy example, and then show relevant metrics for a more serious application to the comparison of English messages parsed with WordNet\cite{wordnet}. Finally, there will be a brief discussion of potential use cases and future outlook.

\section{Proposed Algorithm}
\label{sec:algo}
\subsection{Encryption}
\label{subsec:encryption}
The algorithm begins by encrypting each of the two sets to be compared. Since the encryption procedure is the same for each set, let us focus on one of them, denoted $W$.

Though $W$ may in general consist of elements of any type (e.g. strings, integers, floats, etc.) we will assume they are drawn from members of a particular master set $X$, i.e.  $W \equiv \{w_1, ~ w_2, ~...,w_N\}$ with $W \subseteq X \equiv \{x_1,~ x_2, ~...,x_P\}, ~~ P \ge N$.

We will also need a map $M$ defined which takes any  $x$ in $X$ to a set of integers $\Omega_{x}$, valued in $R \equiv [0,I_{max}]$, for some large positive integer $I_{max}$, i.e. $M: x \to \Omega_{x}$.
In particular,
\begin{eqnarray*}
\label{eqn:map}
M: ~~ w_i & \longrightarrow & \Omega_{i} \\
    & \equiv & \{\Omega_{i1},~ \Omega_{i2},~ ... , \Omega_{ip_i} \}
\end{eqnarray*}
with $\Omega_{ip}$ being the p-th element of $\Omega_i$.
Then the encryption $S_n$ of $W$, for a configurable parameter $n$ that controls the strength of the encryption,  is defined as the set

\begin{equation}
\label{eqn:sn}
S_n \equiv  \{\Omega_{i_{1}p_{1}} + \Omega_{i_{2}p_{2}} + ... + \Omega_{i_{n}p_{n}}\}_{1 \le i_1 < i_2 < ... < i_n \le N; ~ p_1,~ p_2, ... p_n}
\end{equation}
with duplicate elements removed (this turns out to be important for storage and security purposes later). This is hereafter known as ``n-Sum Encryption". Note that for $S_n$ to be defined we require $W$ to have at least $n$ elements, i.e. $N \ge n$.

The simplest variant, $n=1$, gives a set $S_1$ consisting of the union of all the $\Omega_{i}$:
\begin{equation}
\label{eqn:s1}
S_1 \equiv \{\Omega_{i}\}_{i\le N}
\end{equation}
For $n=2$ the encryption set is the set of all sums of pairs of integers, each originating from a different $\Omega_{i}$:
 \begin{equation}
\label{eqn:s2}
S_2 \equiv \{\Omega_{ip} + \Omega_{jq}\}_{i < j \le N; ~ p,~q}
\end{equation}
For $n=3$ all triplet combinations must be computed:
 \begin{equation}
\label{eqn:s3}
S_3 \equiv \{\Omega_{ip} + \Omega_{jq} + \Omega_{kr}\}_{i < j < k\le N; ~ p,~q,~r}
\end{equation}

This pattern extends to higher values of $n$ where one computes quartets, quintets and so on.
For $n > 1$ we have a legitimate encryption procedure because it is much easier to sum several numbers than to uniquely reconstruct them from that sum, this being essentially the NP-hard ``subset sum problem" (Appendix A contains an explicit Python script to perform the encryption which we will make use of later in Section \ref{sec:messages}).

Breaking the encryption, i.e. inferring $W$ from $S_n$ given knowledge of $M$, requires brute-force search\footnote{As noted, $S_1$ is not very secure as it suffices to walk through each integer $x$ in $R$ and test $\Omega_{x}$ for inclusion in $S_1$. This can be done in linear $O(I_{max})$ time.} and depends strongly on $n$ and $\Omega$.\footnote{For example, in the degenerate case where $\Omega_{i} \equiv \Omega~~\forall i$, we always have $S_1 = \Omega$ for any input set $W$, thus we can infer nothing about it. The other extreme is where $\Omega_{i} \bigcap \Omega_{j} = 0~\forall i,j$, and the information of $S_n$ is maximized. Indeed here we can uniquely reconstruct $W$ from $S_1$.} For $S_2$, for example, brute force decryption requires iterating over all pairs $(x,y)$ in $R$ and testing for inclusion of $\{\Omega_{xp} + \Omega_{yq}\}_{p,q}$ in $S_2$, which has quadratic $O(I_{max}^2)$ time complexity. In addition, a certain combinatorial uncertainty arises in this procedure, for $\{\Omega_{xp} + \Omega_{yq}\}_{p,q}$ may be included in $S_2$ without either $\Omega_{xp}$ or $\Omega_{yq}$ being included in any of the $\Omega_i$ mapped from $W$, i.e. one can expect false positives which are in general irreducible by the subset sum problem, for knowing a sum of two or more integers does not in general uniquely identify those integers, even if one knows the set from which they are drawn.

  In general, brute force decryption of $S_n$ would require both searching the space of all n-tuplets in $R$, taking $O(I_{max}^n)$ time\footnote{the number of terms contributing to $S_n$ increases like $^{N}_{n}C \approx \frac{N^n}{n!}$.}, as well as dealing with worsening combinatorial uncertainty; already for $n \ge 3$ and $I_{max} > 10^8$ commodity hardware would really struggle to search n-tuple space (which here is the size of Avogadro's number!), and then only to be swamped with false positives.

\subsection{Finding the Intersection}
\label{subsec:intersect}
The goal of PSI is for the two users to discover the intersection of their respective private sets $W$ and $W'$. To do this they simply measure the intersection of their encrypted sets $S_n$ and $S_n'$: by construction if both private sets share at least $n$ elements in common, even partially, this intersection will be non-empty and each user can easily find which n-tuple(s) are responsible for that intersection, and to what extent. In the process of computing $S_n$, for example, the user can build an inverted dictionary mapping elements of $S_n$ to lists of elements from the $\{\Omega_i\}$. Then knowing $S_n \bigcap S_n'$ identifies a subset of $S_n$ which the dictionary then maps to a set of elements $Y$, and we can assign a `matching score' to each of the original elements $w_i$ given by the ratio $\frac{|Y \bigcap \Omega_i|}{|\Omega_i|}$.

\section{A Geometrical Application}
\label{sec:toy}

\begin{figure}[!htb]
\begin{center}
\includegraphics[width=6in]{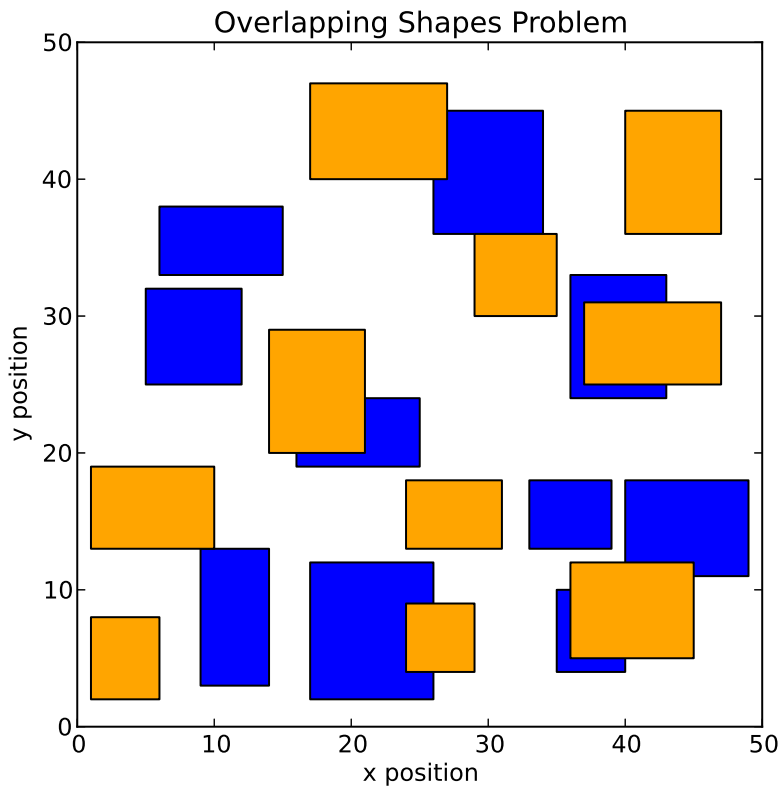}
\end{center}
\caption{\small \emph{Example of the overlapping shapes problem. Orange and Blue wish to discover their mutual intersection without disclosing anything else about the sizes and positions of their shapes. In this example there are $N=10$ Blue and Orange colonies each with random widths and heights between 5 and 10 (arb. units), territory size = $50 \times 50$; coordinates are hashed to random integers between 1 and $I_{max} = 10^8$ with flat  priors.}
}
 \label{fig:shapes}
\end{figure}

As a toy example, consider the problem of privately determining the geometric intersection of two groups of shapes (see Figure \ref{fig:shapes}, and Appendix B for generating code), assumed rectangular for simplicity though the discussion below will apply just as well to any shape. This might represent two countries (`Blue' and `Orange') which each want to secretly establish a certain number colonies of various sizes in a given territory without overlapping with each other, as such overlapping is presumed to be mutually disadvantageous. It's in the best interest of each country to know what the potential conflict areas are, but neither is willing to disclose any other information about the locations and sizes of their colonies. Here the private sets are the respective colonies ($w_i$, $1 \le i \le N$), each of which is mapped to a collection of grid points ($\Omega_i$), suitably hashed\footnote{There are several choices here: the trivial hash, i.e. simply numbering coordinate points from 1 to M (= number of grid points in the territory), a flat random hash assigning each coordinate to a random number between 1 and $I_{max}$, or a custom hash taking coordinates to some values with a given distribution. The trivial hash gives poor results because of a low $I_{max}$, e.g. for $S_2$ all bi-sums are between $0$ and $2M$ and most are degenerate at the peak of a pyramid-shaped distribution; one does better for a flat random hash with a much larger choice of $I_{max}$, or a custom hash designed such that the distribution of hash values results in a flat distribution of bi-sums.}, that compose the rectangular areas. For example, a colony centrally located at $(10,10)$ with a width and height of 3 might be mapped to 9 numbers representing the enclosed coordinate points:
 \begin{eqnarray*}
  \Omega_{\{(10,10),3,3\}}  &=& \{a_1, a_2, a_3, a_4, a_5, a_6, a_7, a_8, a_9 \} \\
\end{eqnarray*}
 while another located at $(12,12)$ with the same width and height would be
  \begin{eqnarray*}
  \Omega_{\{(12,12),3,3\}}  &=& \{a_9, a_{10}, a_{11}, a_{12}, a_{13}, a_{14}, a_{15}, a_{16}, a_{17} \} \\
\end{eqnarray*}
overlapping at exactly one point ($a_{9}$) as expected.
With each country agreeing to the same standard map $M$: colony $\to \Omega$, they can proceed to compute their respective $S_n$ and compare these sets to discover their planned colonies' intersection.

\begin{figure}[!htb]
\begin{center}
\includegraphics[width=2.5 in]{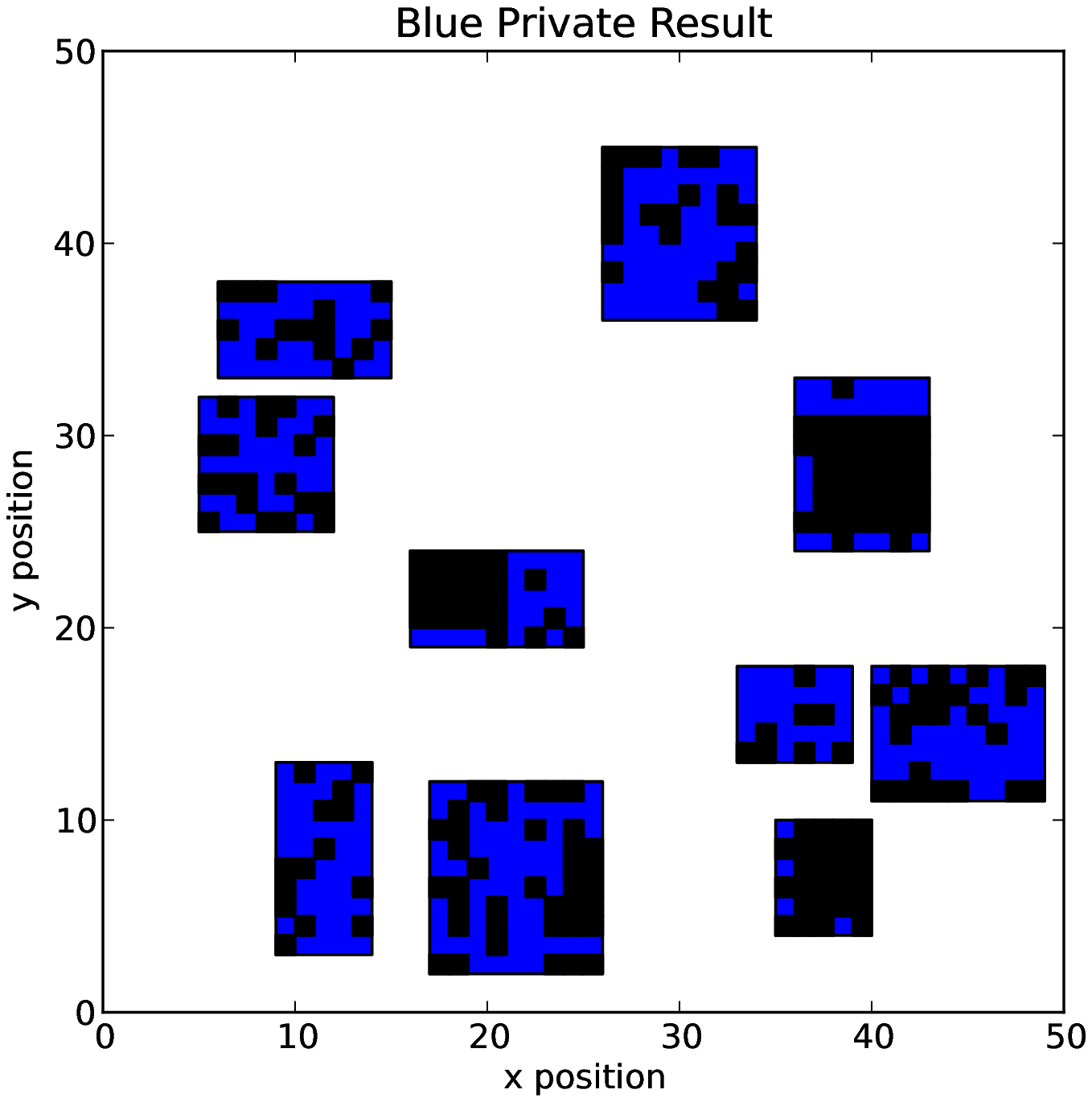}
\includegraphics[width=2.5 in]{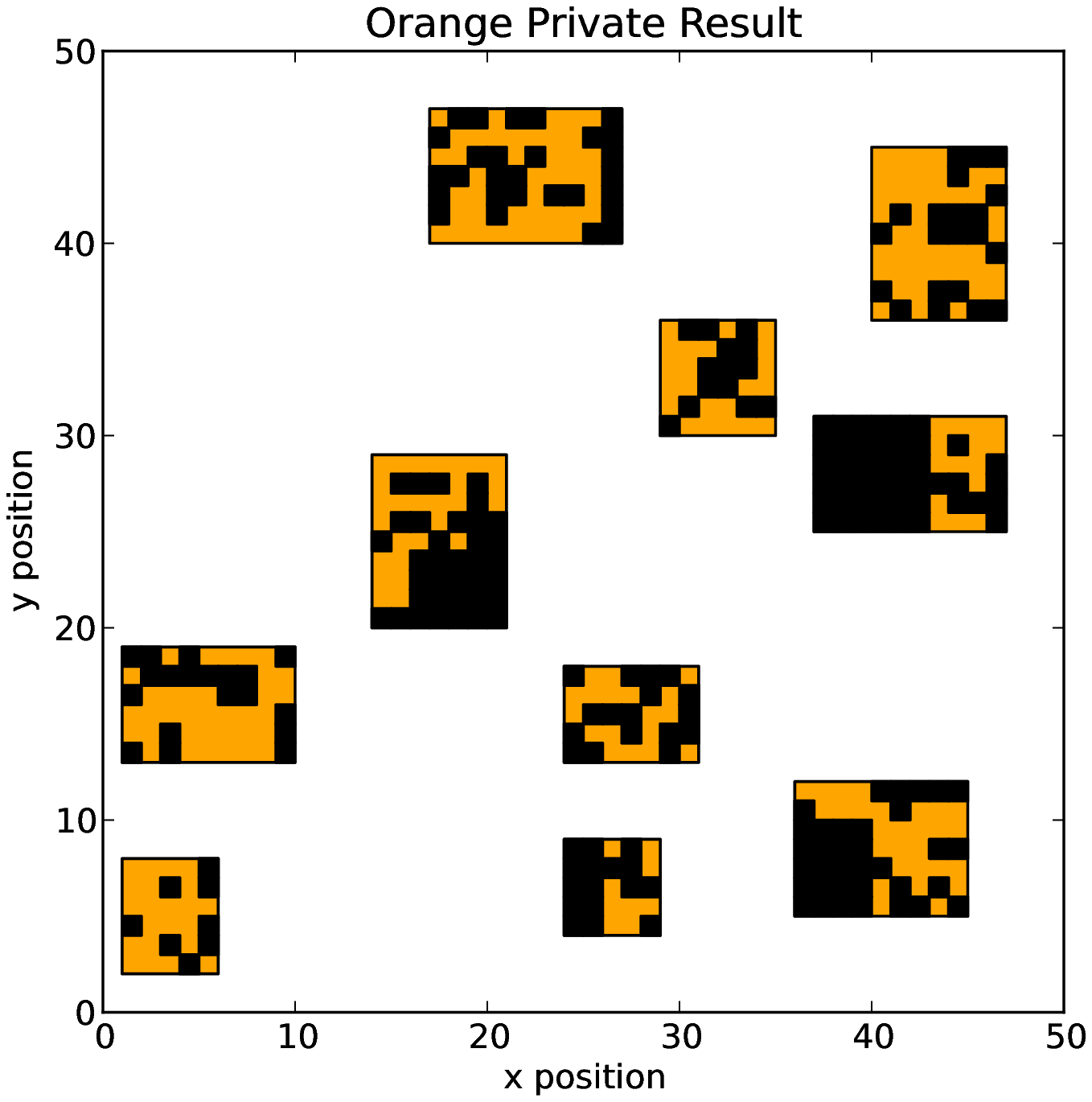}
\includegraphics[width=2.5 in]{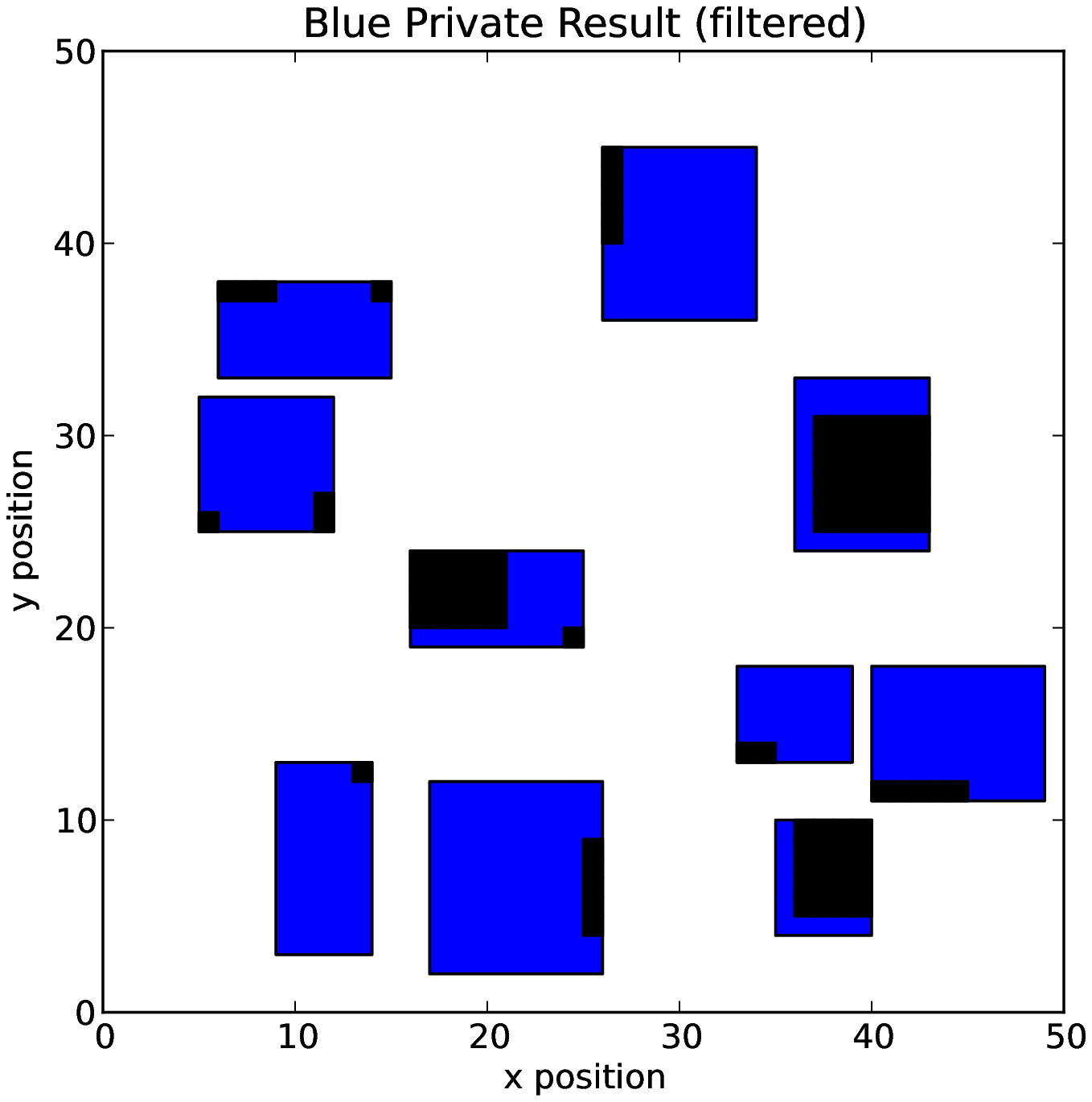}
\includegraphics[width=2.5 in]{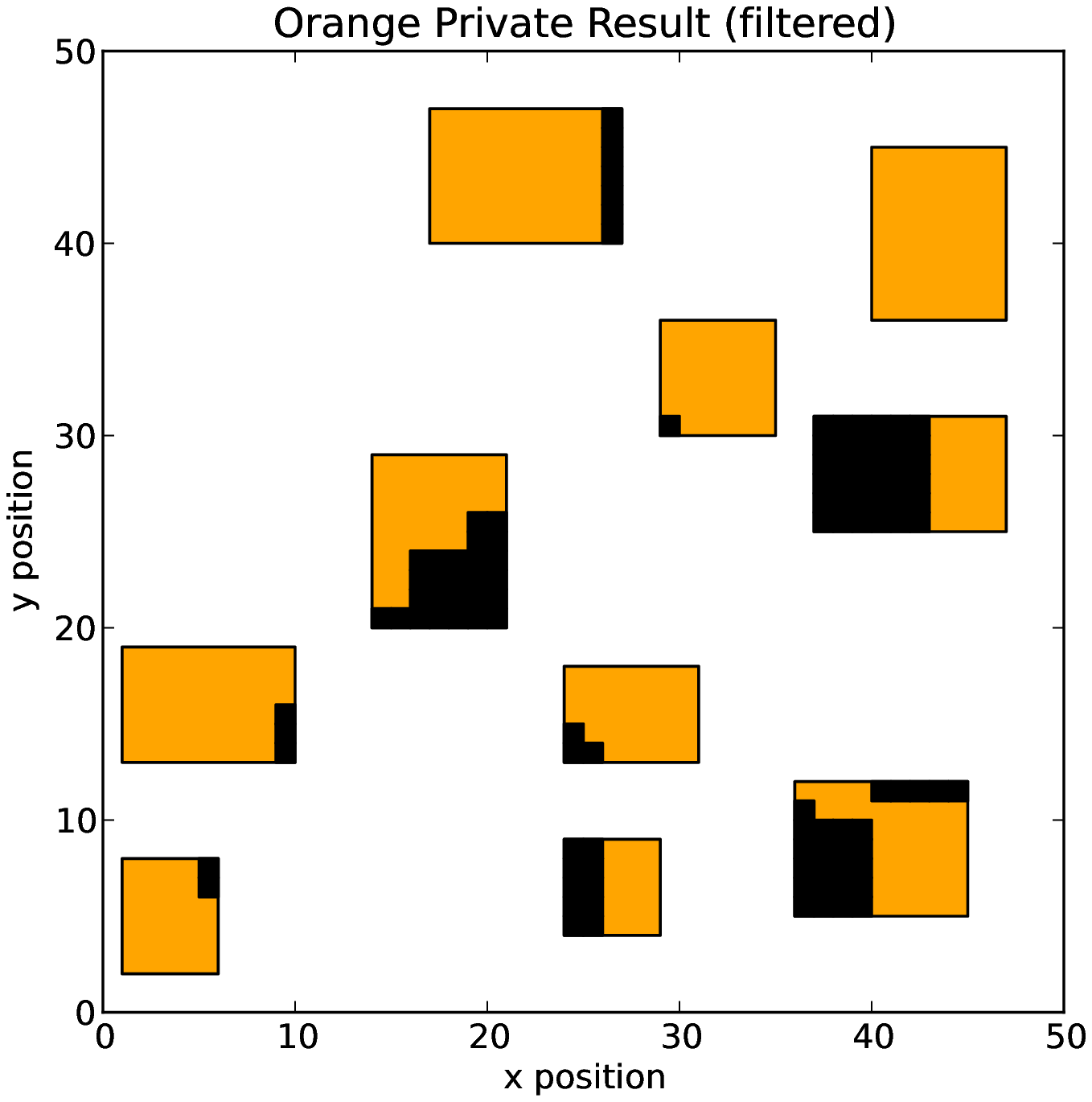}
\end{center}
\caption{\small \emph{Blue (LEFT) and Orange (RIGHT) views of the world: candidate overlap points (black) based on raw comparison of encrypted data. After geometric filtering (BOTTOM), the candidate overlap regions are cleaner and agree almost perfectly with the actual intersection (see previous Figure).}
}
 \label{fig:clean}
\end{figure}

For definiteness let the countries compare $S_2$ computed over their respective colonies. It is now a straightforward procedure for each to map the intersection back to the original points (encoded in the $\Omega_i$) to determine which of these led to the intersection: the upper plots of Figure \ref{fig:clean} show these candidate points in black. There are false positives in this procedure forming a certain combinatorial background, or `fuzziness',  but no false negatives: if a point is not black, it definitely is not in the actual intersection of the colonies. This stage of the comparison may already be useful if Orange and Blue simply want to know which parts of their planned colonies would definitely not be overlapping. But we can do better, for in this case we have a geometrical way of filtering out most of the false positives: since we know the colonies are rectangular (and say of a given minimum size, here taken as 5 grid units), candidate points must compose rectangular regions of at least that minimum size. After this filtering, each country obtains a much cleaner representation of the overlap without learning anything else about the other country's colonies (lower plots of Figure \ref{fig:clean}), which by comparison to Figure \ref{fig:shapes} the reader can verify is quite accurate. If we wanted to report a matching score for each shape, that would simply be given by the ratio of black to non-black area.

\begin{figure}[!htb]
\begin{center}
\includegraphics[width=6in]{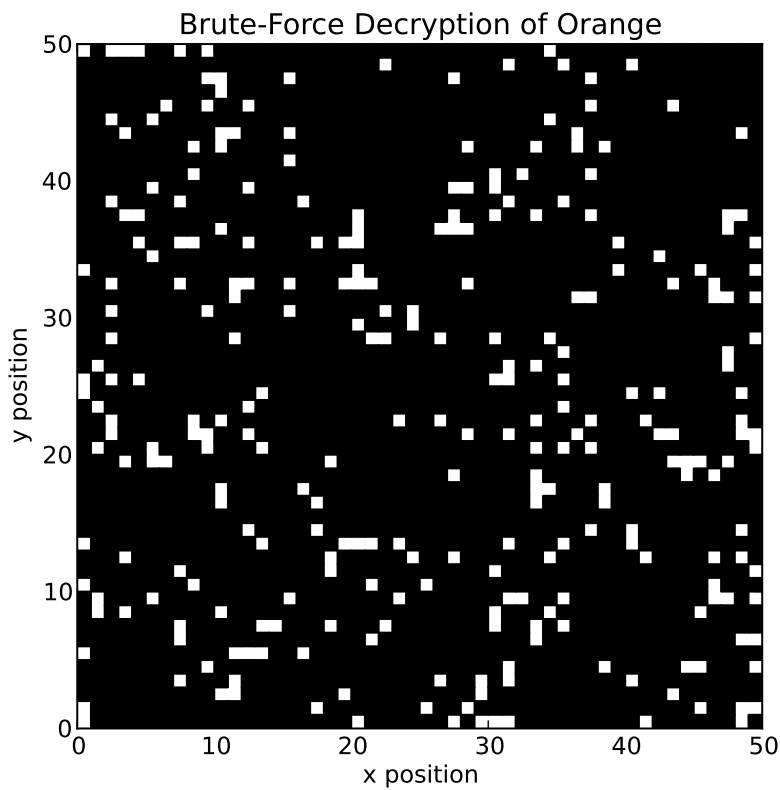}
\end{center}
\caption{\small \emph{Result of brute-force decryption of Orange's $S_2$ data. Shape data is sufficiently obfuscated.}
}
 \label{fig:brute}
\end{figure}

Each country's $S_2$ is safe against brute-force decryption, which would involve not only generating a look-up table of encrypted values to points\footnote{In this toy example this is actually easy, as there are only $M = 50 \times 50 = 2500$ possible hashed grid points, thus in principle $^M _2 C$ = 3 million possible sums (rows in the table).}, but then trying to back-solve to find all the candidate points responsible for the observed $S_2$. As long as the maximum hash value $I_{max}$ is sufficiently small the vast majority of the sample space will be candidate points and very little information is gleaned (see Figure \ref{fig:brute}). Note that if $I_{max}$ is too small then the matching procedure between the countries likewise becomes uninformative (by the Central Limit Theorem sums between random integers tend to cluster about a central value\footnote{But see a prior footnote about choice of distribution.}), so for a given setup there is an optimal value that provides good discrimination for matching but poor discrimination for brute-force decryption; for the parameters of the present toy model that balance is struck at roughly $I_{max} = 10^8$.

\section{Application to Message Comparison}
\label{sec:messages}
The motivating application is the comparison of text messages. Like the colonies of the toy example above, words are intrinsically extended objects that can overlap with each other, but in semantic space instead of geometric space, e.g. the word ``fast" shares some semantic overlap with ``quick", ``hasty", and ``instant", all in differing amounts, and WordNet provides the map from a word $w_i$ to its ``semantic territory" $\Omega_i$ that allows us to compute that overlap. Specifically, each $\Omega_i$ is a collection of synonym sets (``synsets" for short) that are represented by a particular positive integer (``offset" in WordNet parlance)\footnote{The actual $\Omega_i$ we use here are computed from the WordNet database as follows: for a given word $w_i$, we first find all the synsets that contain that word, and for each of these we look up all of their associated synsets, excepting those of type ``antonynm". Thus $\Omega_i$ includes all synsets up to two links removed from $w_i$.}; they are the result of extensive linguistic research and generally ``make sense", i.e. given any three words x, y, and z,  if x is semantically closer to y than to z, then typically $|\Omega_x \cap \Omega_{y}|  > |\Omega_x \cap \Omega_{z}|$. It is useful to keep the analogy to the geometrical application of the last section in mind with the correspondences of country $\to$ message,  colony $\to$ word, and coordinate $\to$ synset integer, for we really have the same problem in a different guise, i.e. we are considering the problem of how two users can compare messages in order to discover their words' semantic intersection only.

Given an English message then, we map the words to the appropriate numerical identifiers and apply the cryptographic hash $S_n$.
Note that applying this encryption to messages is implicitly using the ``Bag of Words" NLP model wherein word order, grammar, and punctuation are ignored. Despite the crudeness of that model, one can reliably match large classes of messages, in particular those whose meaning is invariant upon permutation of their words, e.g. ``withdraw money tomorrow noon", ``agriculture thesis topic", etc. In fact the set $S_n$ applied to the whole message is equivalent to $S_n$ applied to all possible n-grams constructed from words in the message.

Let us consider a simple, if somewhat contrived example that we can compute by hand:

\bigskip

Message 1:  ``Laser reheat cappuccino"

\bigskip

Message 2: ``Laser reheat espresso"

\bigskip
Each word belongs to a particular synset,

\begin{eqnarray*}
``laser" = w_1 &\equiv&   3643253 \\
 ``reheat" = w_2 &\equiv&   544280 \\
 ``cappuccino" = w_3 &\equiv&   7920349\\
 ``espresso" = w_4 &\equiv&  7920052\\
\end{eqnarray*}
and we look up the mapping to other synsets in WordNet:
{\small{
\begin{eqnarray*}
  \Omega_1 &=& \{\textbf{3643253}, \textbf{3851341}, \textbf{3924532}\}\\
 \Omega_2 &=& \{\textbf{371264}, \textbf{544280}\}\\
 \Omega_3 &=& \{7920349, \textbf{7929519}\}\\
 \Omega_4 &=& \{7920052, 7920222, \textbf{7929519}\}\\
\end{eqnarray*}
}
}
the bold-faced synsets being common to both messages.
Following (\ref{eqn:s2}), we have what we need to compute $S_2$ for each message:
{\small{
\begin{eqnarray*}
  S_2 &=& \{4014517,4187533,4222605,4295796,4395621,4468812,8291613,8300783, \\ &&
  8464629, 8473799,11563602,11572772,11771690,11780860,11844881,11854051\}\\
  \\
 S_2' &=& \{4014517,4187533,4222605,4295796,4395621,4468812,8291316,8291486, \\ && 8300783,8464332,8464502,8473799,11563305,11563475,11572772,11771393, \\ && 11771563,11780860,11844584,11844754,11854051\}\\
 \\
 S_2 \bigcap S_2' &=& \{4014517,4187533,4222605,4295796,4395621,4468812,8300783,
   8473799, \\ && 11572772,11780860,11854051\}\\
\end{eqnarray*}
}
}

Here $S_2$ has 16 keys, $S_2'$ has 21, and $S_2 \bigcap S_2'$ has 11, giving each message an ``overlap" percentage of $11/16 \approx 69\%$ and $11/21 \approx 52\%$, respectively. In this simple example each user can easily determine exactly which synsets in their respective $\Omega_i$ contributed to that overlap (the bold values) by keeping track with an inverted dictionary; in this case there is no uncertainty, and thus we say the ``confidence" of matching is $100\%$.

Brute-force decryption of $S_2$ would again require not only the costly construction of a look-up table\footnote{In this case for $S_2$ this is still doable, barely: there are $M \approx 1.5 \cdot 10^5$ words in WordNet3.1, thus in principle $^M _2 C$ = 10 billion possible sums to tabulate; this might fit in the memory of a commodity server. Doing the same for $S_3$ and beyond would quickly become impractical, however.} but also some way to deal with the inherent combinatorial uncertainty of this process. For example, supposing we wished to brute-force decrypt the more embellished message ``US car production blue sapphire laser millisecond pulse reheat", which has 62461 keys: in addition to the message's actual words, it turns out the brute-force decryption also returns hundreds of red herring match results for words like ``no-go", ``keyless", ``brainwashed", and ``baptised". This is analogous to what brute-force decryption did for us in the geometrical example of the last section, i.e return the almost non-informational Figure \ref{fig:brute} which basically says that most points are \emph{possibly} in the intersection.

\begin{figure}[!htb]
\begin{center}
\includegraphics[width=6in]{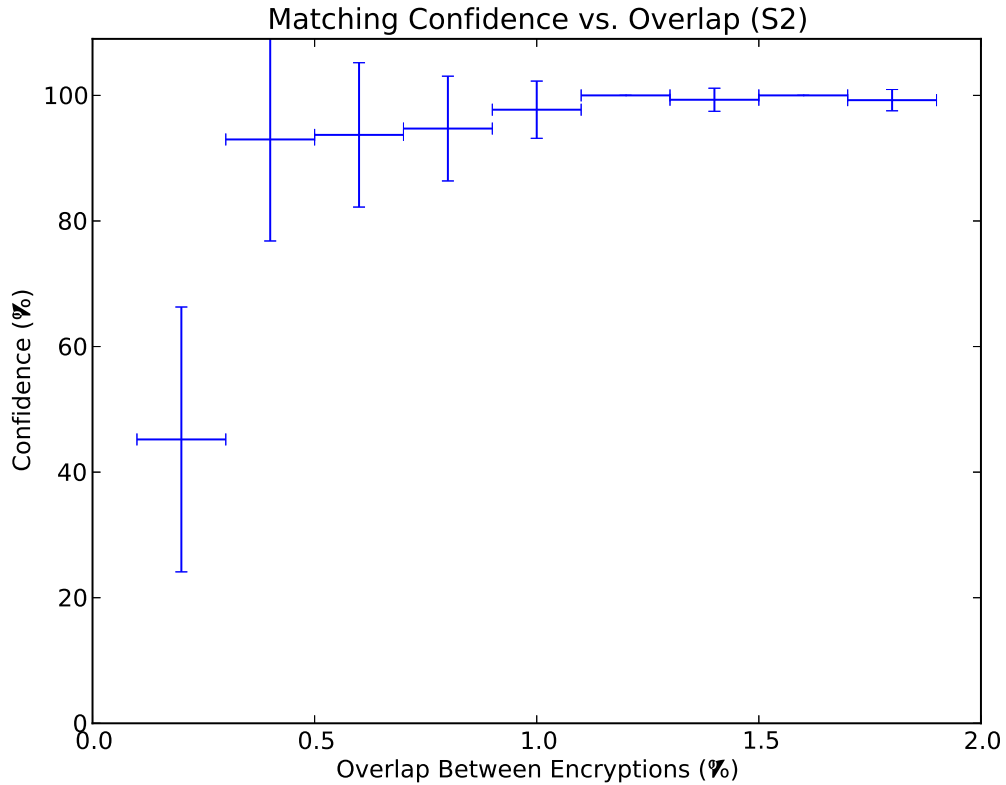}
\end{center}
\caption{\small \emph{The number of keys in the overlap relative to the total number is an indicator of the cleanliness of the process: for a generic 10-word message encrypted with $S_2$, an overlap in excess of 1\% of the total number of keys essentially guarantees that any synset leading to the overlap is a true positive.}
}
 \label{fig:confidence}
\end{figure}

In the toy geometrical example we learned that the matching procedure produces many false positives in addition to real positives, and that there was a simple way to filter out the false positives based on geometrical constraints on the way the rectangular shapes can overlap. Unfortunately in the present case there is no such obvious filter we can apply: the synsets that represent a word are not localized into any special ``shapes" in a way that we can leverage, so we're stuck with false positives in this procedure\footnote{Note we could redefine the synset numbers to give a more even sum distribution, e.g. to optimize for $S_2$ we choose the distribution of synset numbers such that the distribution of bi-sums is approximately flat. As the distribution of synset numbers in WordNet is approximately flat, i.e. $\rho(x) \sim 1$, there is potential for improvement if we redistributed according to something like  $\rho(x) \sim \frac{1}{\sqrt{x}} + \frac{1}{\sqrt{I_{max} - x}}$.}. However, it does turn out that the number of positives (both real and false) is inversely correlated with the number of false positives. For example, measurement of hundreds of randomly-generated pairs of 10-word messages encrypted with $S_2$ (see Figure \ref{fig:confidence}) indicates that an overlap between messages above 1\%  essentially guarantees an absence of false positives, hence a high degree of confidence in the synsets extracted from this procedure. This is in tune with our expectations for a system with inherent uncertainty: the larger signal (overlap) available, the higher one's confidence in conclusions based on that signal.

\subsection{Encryption Storage}

\begin{figure}[!htb]
\begin{center}
\includegraphics[width=6in]{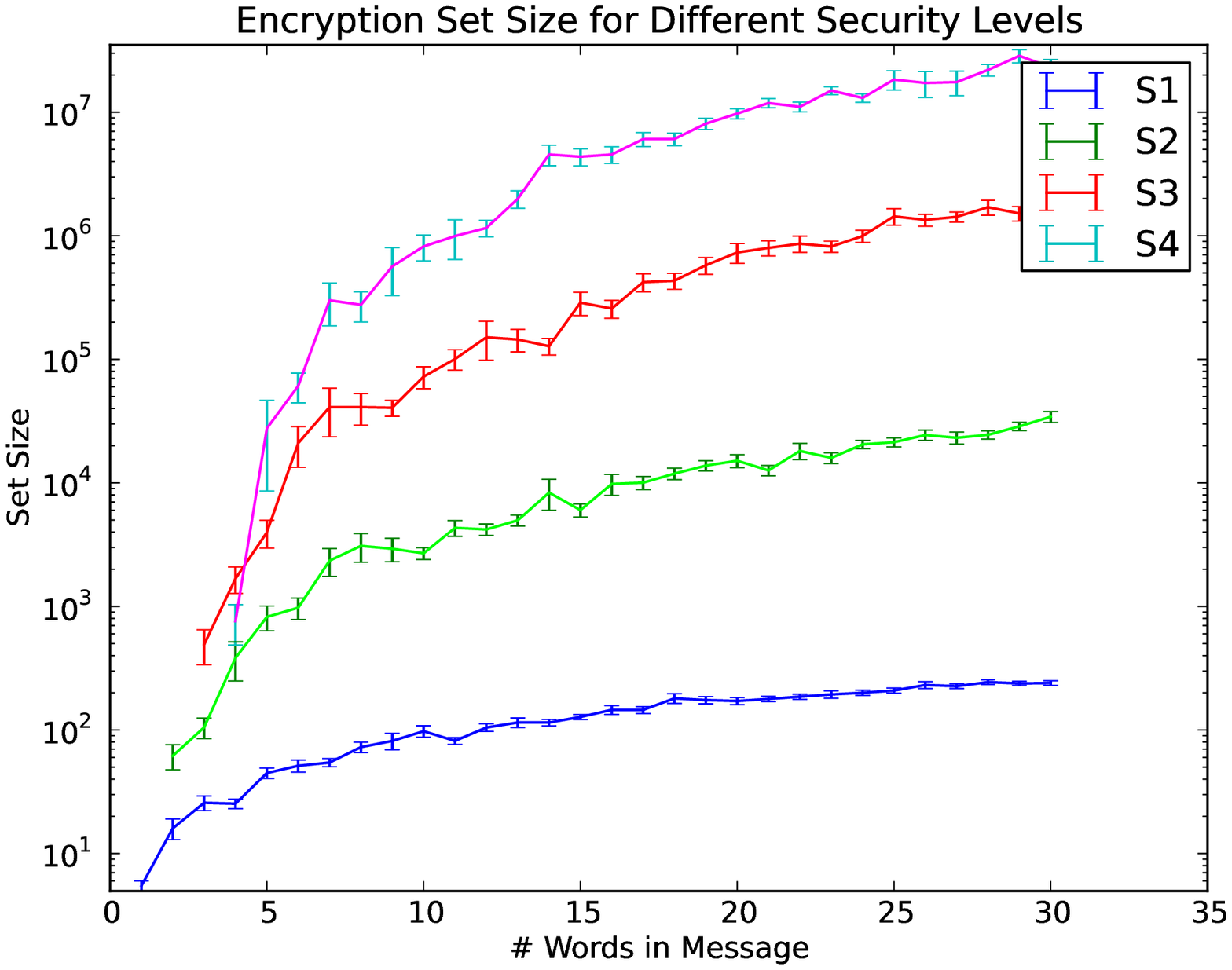}
\end{center}
\caption{\small \emph{Average encrypted set size for private input messages with various numbers of words for encryption methods $S_1$, $S_2$, $S_3$, and $S_4$. Each point is the statistical average of 100 messages whose words were randomly selected from the WordNet 3.1 database.}
}
 \label{fig:random}
\end{figure}

The space required to store an encryption $S_n$ is not entirely negligible. For average-length messages encrypted with $S_2$ one typically produces a list of thousands of keys (i.e 10-100 KB), and this increases exponentially at higher encryption levels (see Figure \ref{fig:random}). The ``US car production blue sapphire laser millisecond pulse reheat" example as a text file is over 500K, though with bit packing the 62461 keys (each a 32-bit int) it really only needs half that space. Standard compression tools can further reduce that by a factor of several\cite{compress}; in this way most encrypted messages typically take the space of a thumbnail image.

\subsection{Encryption Speed}

 Encryption of a set only has to occur once in this algorithm, and this constitutes the bulk of the overall time spent. Comparison to other sets then follows intersection algorithms on sorted sets, which are fast: sub-ms for small ($< 10^4$ elements) to 10~ms for large sets with millions of elements\cite{intersection}. Figure \ref{fig:times} below shows upper-bound timing metrics for encryption -- note these are non-optimized single-threaded speeds, whereas the algorithm is in fact highly parallelizable (e.g. each n-tuple combination in (\ref{eqn:sn}) can run on its own thread), so these times could easily be an order of magnitude smaller in a production environment.\footnote{Encryption metrics are reported here with non-optimized Python code, show in Appendix A, not only for both human readability and portability, but also to deliberately produce pessimistic results.}   For all but the highest levels of encryption ($S_4$ and higher) this initial runtime overhead is comparable to what current PSI techniques use \emph{in every set comparison}. For situations where one set will be compared multiple times to other sets, this overhead is amortized over the number of comparisons performed and we thus quickly attain a speed limited only by ordinary set comparison, which as mentioned is at least two orders of magnitude faster than current PSI techniques.

\begin{figure}[!htb]
\begin{center}
\includegraphics[width=6in]{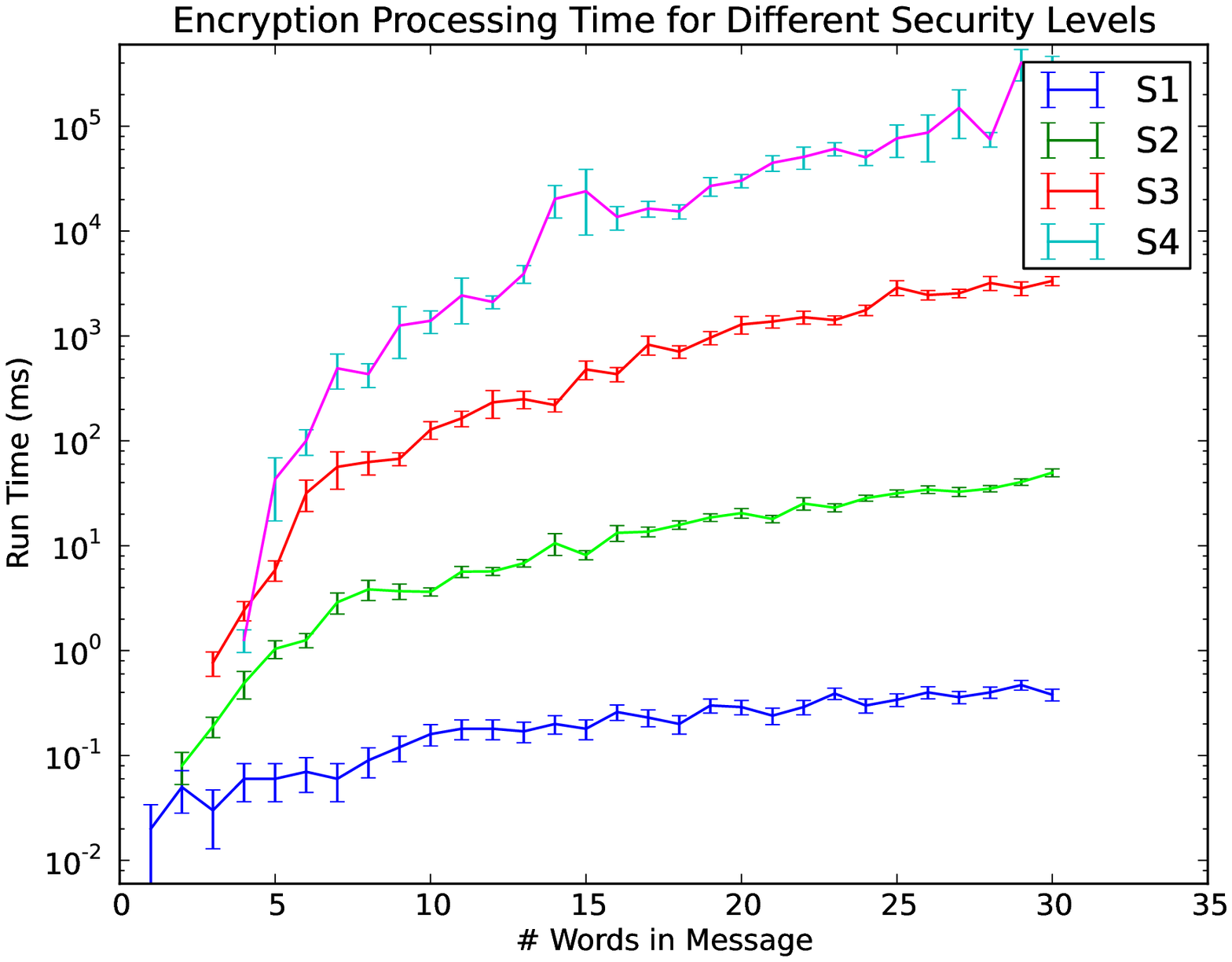}
\end{center}
\caption{\small \emph{Average processing time required for encryption of private input messages generated for the previous figure. Encryption code was executed single-threaded and non-optimized using the Python 2.6.7 compiler on a commodity laptop (Intel i7-2640M CPU @ 2.80GHz, 4.00 GB RAM). }
}
 \label{fig:times}
\end{figure}

\section{Use Cases for Fast PSI}
\label{sec:uses}
Private Set Intersection techniques promise to enjoy a large number of potential applications, but these are made more feasible by a fast protocol such as the one proposed. In the social domain, i.e. interest matchmaking,  it is quite likely that users will offer a much more liberal set of interests than they currently do with public protocols owing to the reassurance that non-matching interests will not be revealed to anyone. The fact that their private lists can be persistently stored in encrypted form removes the need for a simultaneous transfer protocol: one-sided matching is a very fast and practical option. A few examples:
 \begin{itemize}
   \item Privacy-Preserving Social Networking: users compare lists of private profile descriptors to make relevant new connections.
   \item Advertising: marketers compare a list of potential features of the ad with consumers' private lists of interests in order to predict the ad's success.
   \item Scientific Collaboration: researchers compare lists of key words to find potential collaborators on a topic without announcing their idea or intentions to the public and risk being scooped.
   \item Employee Complaint Forum: employees compare lists of concerns to determine whether a threshold commonality of complaints is reached, prompting management for reform with minimal danger of incrimination.
   \item Social Organizing: individuals compare lists of place+time+topic in order to effect a spontaneous social gathering.
   \item Meeting Agendas:  participants submit private lists of desired topics, and the overlap is used to draw up an agenda.
   \item Corporate Benchmarking: companies compare confidential lists of assets to determine ranking (a sort of generalization of Yao's ``Millionaire Problem"\cite{yao}).
 \end{itemize}

 On larger organizational or governmental scales there are many more promising applications, e.g.
 \begin{itemize}
   \item Medicine: patients' medical records are compared to a specific `filter list' in order to identify and process records without accessing them in entirety.
   \item Law Enforcement: a federal agency needs to compare its confidential list of suspects with some other institution's likewise confidential list of registrants.
   \item Personalization: multi-component systems can compare lists of users' attributes to certain criteria in order to identify recurrent users and provide a personalized experience.
   \item Routing Optimization: network nodes compare lists of private attributes such as capacity and flow constraints in order to determine a globally optimal routing.
   \item Central Command Processing: a central unit can publicly broadcast a single command (encrypted list of goal keywords) which worker-units compare with their own private list of qualifications in order to effect the appropriate action, maximizing communication efficiency and security.
 \end{itemize}

\section{Conclusions and Remarks}
\label{sec:conclusions}

This paper introduces a new Private Set Intersection technique using n-Sum encryption that should be many times faster than current techniques when any given set is likely to undergo multiple comparisons in its lifetime. The speed increase is due to the fact that, after an initial one-time encryption of each set, all subsequent comparisons proceed by ordinary comparison of ordered sets, and many well-studied fast algorithms exist for this.

The technique is mathematically simple and flexible in that one has a choice of which mapping of set elements $M: x \to \Omega$ to use. Specifying the technique in terms of an abstract map $M$ has the advantage of decoupling: different choices of $M$, be they versions of WordNet or customized hashes, can be tailored to the application in hand. In general $M$ is chosen to make reporting of partial intersection between set elements feasible and meaningful, with the traditional binary `yes/no' reporting a result of choosing a trivial $\Omega$, i.e. $M: ~x \to x$.
As this is mostly a proof-of-concept paper, the technique is likely to be highly optimizable, leveraging especially the highly-parallelizable nature of the algorithm, and adaptable to different technologies.

Security in n-Sum encryption is configurable: while $S_1$ uncovers the bare content of the sets, $S_2$ and $S_3$ offer security at a level which is probably sufficient for casual PSI, while $S_4$ and $S_5$ may be suitable for more sensitive (e.g. financial and governmental) applications of PSI. Malicious brute-force decryption, or equivalently submitting a huge set of every possible element to troll for matches, doesn't pay: not only does the computational burden become formidable for $S_3$ and above, but  a huge set intersected with a small one has a very low percentage-wise overlap, so (e.g. Figure \ref{fig:confidence}) the intersection will contain mostly false positives. The most effective matching is expected to be a small submission against another small submission, so this technique is ideal for finding rare matches between users in a large pool that otherwise wouldn't easily find each other.

Uncertainty or `fuzzy matching' is a side effect which may actually be advantageous. Oftentimes communication does not require exact clarity, but rather approximate accuracy on the first pass.\footnote{`Hearsay' is also a fuzzy mode of communication which enjoys the features of being certain enough to be highly useful in daily life, but too uncertain to be used in court to reap punishment. It is safe, cheap talk. Such modes of communication probably naturally evolve in systems to optimize collective results\cite{cheap}.} Furthermore, as we saw above fuzziness makes brute-force decryption much harder to interpret.

Finally, one can imagine various usage scenarios: in one scenario there is a public file server where one can freely upload and download (encrypted) sets\cite{qr}; the server thus might maintain a directory of all submissions, a mix of sets encrypted with different security levels and different $M$, perhaps, and anyone can intersect with their own set and initiate an exchange. At the opposite extreme, a closed boardroom-type scenario would utilize a small private server to collect and compare submissions from invited participants. Also, in the scenario of a location-based smartphone app, one might initiate a Bluetooth exchange with other nearby devices subject to certain constraints, e.g. set size, security level, user identity/trustworthiness, and of course proximity.

\section*{Appendix A: n-Sum Encryption}

The following runs in Python 2.6.7

\small{
\begin{verbatim}
# N-Sum encryption for any N
#  BEFORE RUNNING: put this script and "omegaO2.txt"
#                 (https://github.com/nkersting/WordNet) in the same directory.
#  USAGE: Simply run this script and, when prompted, enter the message
#         you wish to encrypt.
#  OUTPUT: Sorted list of encryption keys written to "keys.txt", one per line
##################################################################
# This script produces usable sum-encrypted messages, but there are still some
# additional needed improvements. For example,
# 1. tokenization, e.g. handle punctuation (note convert "read." to "read"
#    but not "U.S." to "U.S")
# 2. lemmatization, e.g. convert plural nouns to singular (WordNet has no plurals)
# 3. hash words not in the dictionary to unused unique integers
#    (WordNet uses integers between 0 and roughly 1.6*10^7)
# 4. compound words (connected with "_")
# 5. Part of Speech differentiation

import csv
import shlex

###################################################################
def MapEntry(entryword, syndict, stopwords):
    if (syndict.has_key(entryword) and entryword not in stopwords):
        return syndict[entryword]
    else:
        return []

###################################################################
def SumWordsDFSRecursive(currwords, index, syndict, stopwords, sublist, sumdict):
    if index == len(currwords):
        total = sum(sublist)         # key computed here
        if (sumdict.has_key(total) == False):
            sumdict[total] = []
        for item in sublist:
            sumdict[total].append(item)  # build the inverted dictionary
        return

    for entry in MapEntry(currwords[index], syndict, stopwords):
        templist = sublist[:]
        templist.append(entry)
        SumWordsDFSRecursive(currwords, index+1, syndict, stopwords, templist,
                             sumdict)
    return

###################################################################
def SumNEncrypt(userwords, currwords, syndict, stopwords, N, sumdict):
    if len(currwords) == N:
        sublist = []
        SumWordsDFSRecursive(currwords, 0, syndict, stopwords, sublist, sumdict)
        return

    for i in range(0, len(userwords)):   # iterate over all n-element combinations
        newwords = currwords[:]
        newwords.append(userwords[i])
        SumNEncrypt(userwords[i+1:], newwords, syndict, stopwords, N, sumdict)
    return
###################################################################
def main():

    N = 2   # change this for the desired security level sN
    stopwords = []    # append words to ignore, such as "a", "the", "it", etc.
    synfile = "omegaO2.txt"    # map file defines M: x -> Omega_x
    outfile = open("keys.txt", "w")   # output as a text file, one key per line

    synReader = csv.reader(open(synfile,'rb'), delimiter=' ')   # read map file
    syns = []
    for synline in synReader:
        syns.append(synline)

    syndict = {}        # convert to a dictionary for faster access
    for entryline in syns:
        syndict[entryline[0]] = [int(x) for x in entryline[1:]]

    usertext = raw_input('Message to encrypt: ')    # user input
    userwords = shlex.split(usertext)

    sumdict = {}
    SumNEncrypt(userwords, [], syndict, stopwords, N, sumdict)  # encryption here

    for key in sorted(sumdict.keys()):   # write output with one sum per line
        outfile.write(str(key) + '\n')


\end{verbatim}}

\section*{Appendix B: Overlapping Shapes Code}

This was used for the toy example in Section \ref{sec:toy}

\small{
\begin{verbatim}

def RandRect(collection, worldDim):
    maxDim = 10
    minDim = 5
    passed = False
    while (not passed):
        passed = True
        x = random.randint(1, worldDim - maxDim)  # ensure colonies inside borders
        y = random.randint(1, worldDim - maxDim)
        width = random.randint(minDim, maxDim)
        height = random.randint(minDim, maxDim)
        for member in collection:
            x2 = member[0]
            y2 = member[1]
            width2 = member[2]
            height2 = member[3]
            if (x + width >= x2 and        # ensure each country's colonies
                y <= y2 + height2 and      # non-self-intersecting
                y + height >= y2 and
                x <= x2 + width2):
                    passed = False

    collection.append((x,y,width,height))
    return

####################################
def MakeHash(worldDim):          # hash sequence to large unused integers
    maxHash = 100000000          # store the reverse hash as well
    topInt = worldDim * worldDim # maxHash must be chosen large enough
    used = set()                 # to make the encryption give useful results,
    posHash = {}                 # but not so large that decryption easily
    revHash = {}                 # betrays the input set

    for i in range(1, topInt + 1):
        passed = False
        while (not passed):
            key = random.randint(1,maxHash)
            if (key not in used):
                used.add(key)
                posHash[i] = key
                revHash[key] = i
                passed = True

    return posHash, revHash

###################################
def MakeSumDict(posHash):    # constructs look-up dictionary
    sumToPoints = {}         # for brute-force decryption
    allkeys = posHash.keys()

    for i in range (0, len(allkeys)):
        for j in range (i, len(allkeys)):
            s = posHash[allkeys[i]] + posHash[allkeys[j]]
            if (sumToPoints.has_key(s) == False):
                sumToPoints[s] = []
            sumToPoints[s].append(allkeys[i])
            sumToPoints[s].append(allkeys[j])

    return sumToPoints
####################################
def MapSumsToPoints(sumlist, sumToPoints):

    candidates = set()
    for s in sumlist:
        for v in sumToPoints[s]:
            candidates.add(v)
    return candidates
###############################
def FillColonyOmega(colonies, positionHash, worldDim):

    omega = {}
    count = 0
    for colony in colonies:
        count += 1
        vals = []
        for i in range(0,colony[2]):
            for j in range(0,colony[3]):
                vals.append(positionHash[colony[0] + i + worldDim*(colony[1] + j)])
        omega[str(count)] = vals

    return omega

##########################################
def PointToCoords(point, worldDim):
    x = point % worldDim
    y = point / worldDim
    return x, y




##################################
def main():

    N = 10   # number of colonies in each collection
    worldDim = 50  # number of grid units on each side of the world

    random.seed()

    usertext = ""
    for i in range (1, N+1):            # fake 'message'
        usertext += (str(i) + ' ')      # describing the colonies 1 .. N
    userwords = shlex.split(usertext)

    # hash from coordinate position to particular random int
    positionHash, reverseHash = MakeHash(worldDim)

    colonies1 = []
    for i in range(0,N):
        RandRect(colonies1, worldDim)

    colonies2 = []
    for i in range(0,N):
        RandRect(colonies2, worldDim)

    omega1 = FillColonyOmega(colonies1, positionHash, worldDim)
    omega2 = FillColonyOmega(colonies2, positionHash, worldDim)

    sumdict1 = {}
    stopwords = []
    SumNEncrypt(userwords, [], omega1, stopwords, 2, sumdict1)

    sumdict2 = {}
    stopwords = []
    SumNEncrypt(userwords, [], omega2, stopwords, 2, sumdict2)

    sumlist1 = set(sumdict1.keys())
    sumlist2 = set(sumdict2.keys())
    overlap = sumlist1.intersection(sumlist2)   # keys in the intersection

    encryptedPoints1 = set()     # true and false positives here
    encryptedPoints2 = set()
    for commCrypt in overlap:
        for comm in sumdict1[commCrypt]:
            encryptedPoints1.add(comm)
        for comm in sumdict2[commCrypt]:
            encryptedPoints2.add(comm)

    true1 = set()              # S_1 for one colony
    for key in omega1.keys():
        for val in omega1[key]:
            true1.add(val)

    true2 = set()
    for key in omega2.keys():
        for val in omega2[key]:
            true2.add(val)

    trueoverlap = true1.intersection(true2)   # these is the actual intersection

    firstCollection = []           # coordinates of apparent intersection
    for point in encryptedPoints1:
        x, y =  PointToCoords(reverseHash[point], worldDim)
        firstCollection.append([x,y,1,1])

    secCollection = []
    for point in encryptedPoints2:
        x, y =  PointToCoords(reverseHash[point], worldDim)
        secCollection.append([x,y,1,1])

    decryptCollection = []       # coordinates of brute-force decrypted intersection
    sumDict = MakeSumDict(positionHash)
    candidates = MapSumsToPoints(sumlist1, sumDict)
    for c in candidates:
        x, y = PointToCoords(c, worldDim)
        decryptCollection.append([x,y,1,1])
        
\end{verbatim}}

\end{document}